\documentclass[numberedappendix]{emulateapj}

\usepackage{epsfig} 

\newcommand{\beq}{\begin{equation}}
\newcommand{\eeq}{\end{equation}}
\newcommand{\beqa}{\begin{eqnarray}}
\newcommand{\eeqa}{\end{eqnarray}}
\newcommand{\om}{\Omega_m}
\newcommand{\omt}{\Omega_T}
\newcommand{\lang}{\langle} 
\newcommand{\rang}{\rangle} 

\shorttitle{Safety in numbers}
\shortauthors{Holz \& Linder}

\begin{document} 

\title{Safety in numbers: Gravitational Lensing Degradation\\ 
of the Luminosity Distance-Redshift Relation} 
\author{Daniel E. Holz}
\affil{Theoretical Division, Los Alamos National Laboratory,
Los Alamos, NM 87545 and\\
Kavli Institute for Cosmological
Physics and Department of Astronomy \& Astrophysics,\\ 
The University of Chicago, Chicago, IL 60637}
\email{deholz@cfcp.uchicago.edu}
\and
\author{Eric V. Linder}
\affil{Physics Division, Lawrence Berkeley Laboratory,
Berkeley, CA 94720} 
\email{evlinder@lbl.gov}

\begin{abstract}
Observation of the expansion history of the Universe allows
exploration of the physical properties and energy density
of the Universe's various constituents. Standardizable candles such as
Type Ia supernovae remain one of the most promising and
robust tools in this endeavor, by allowing for a direct measure of
the luminosity distance-redshift curve, and thereby producing
detailed studies of the dark energy responsible for the
Universe's currently accelerating expansion.  As such 
observations are pushed to higher redshifts, the observed
flux is increasingly affected by gravitational lensing
magnification due to intervening structure along the
line-of-sight. 
We simulate and analyze the non-Gaussian probability distribution
function of de/amplification due to lensing of standard
candles, quantify the effect of a convolution over many
independent sources (which acts to restore the intrinsic average
(unlensed) luminosity due to flux conservation), and
compute the additional uncertainty due to lensing on derived
cosmological parameters.
For example, the ``degradation factor'' due to lensing is a 
factor of three reduction in the effective number of usable supernovae
at $z=1.5$ (for sources with intrinsic flux dispersion of 10\%).
We also derive a useful expression for the
effective increased dispersion in standard candles due to
lensing, as a function of redshift.
\end{abstract} 



\section{Introduction} \label{sec.intro}

The next generation of cosmological probes will explore
properties of our Universe with unprecedented precision.
The broad aim is to characterize the global nature of the
cosmology through a mapping of the expansion history of the
Universe, $a(t)$.  This history is fully encapsulated in the
luminosity distance-redshift relation, which depends on 
global quantities such as the
dimensionless dark matter density, $\Omega_m$, dark energy
density, $\Omega_w$, and parameters describing the precise
nature of the dark energy (e.g.~its equation of state). 
Measurements of the expansion history
thus enable determinations of these cosmological parameters,
and in particular shed light on the nature of the dark
energy.

Type Ia supernovae (SNe) are an essential tool in the direct
mapping of the 
subtle changes in the distance-redshift relation due to the
dark energy.
Indeed, observations of supernovae
play a major role in our confidence
that the current expansion of the Universe
is accelerating \citep{riess98,perl99,tonry03,knop03,riess04}. 
These
(calibratable) standard candles \citep{phillips93,rpk95,wgap03} give a
particularly direct and simple translation from  
observable to expansion history: the redshift of a SN is 
a direct measure of the expansion factor $a$ at emission, while the flux
(or magnitude) of the SN relates to its distance
through a cosmological inverse square law, and hence to the
lookback time $t$ to the SN explosion.  Thus an
observation of a SN is
very close to pure cosmography: drawing a map of the
geometry of spacetime.

We have strong indications from cosmic microwave background
radiation measurements (in concert with large-scale structure 
measurements) that the geometry of {\em space} is
flat \citep{wmap03}, corresponding to a critical density
Universe with total dimensionless energy density $\Omega=1$.
This global statement breaks down on scales small compared
with the cosmic (Hubble) scale, where a wealth of
structure, from galaxies to clusters of galaxies, is
apparent (and holds encoded within it a separate brand of
cosmological information).

Large-scale structure also affects our goal of mapping 
the full {\em spacetime} geometry, representing the
expansion history of the Universe. 
This structure, through its gravitational potential, alters the 
propagation of light from distant sources, including standard candles, 
changing the received flux  
and hence compromising our measures of distance. Since structure on 
various scales fills the Universe, {\em every} light ray is affected 
at some level by this gravitational lensing.  However, since
photons are conserved by this lensing, the {\em mean} flux
over the sources is preserved. 
Gravitational lensing therefore adds a statistical
noise to the cosmographic mapping, one that can 
be overcome by averaging over sufficient numbers of distant
sources.

\medskip

In~\S\ref{sec.updf} we discuss the properties of the lensing flux 
magnification 
distribution, and in~\S\ref{sec.convolve} we show how this affects the 
distance-redshift relation.  We carry this through in~\S\ref{sec.cospar} 
to the influence on extraction of cosmological parameters,
yielding both a degradation 
in precision and a bias, and derive a prescription for the
number of standard candles as a function of redshift required to provide averaging adequate 
to dilute the lensing degradation below a desired level.  
The conclusion in~\S\ref{sec.concl} summarizes the circumstances 
in which lensing has a significant impact on cosmological parameter determination. 
The Appendices discuss the 
non-Gaussian effects of translating between flux and magnitude 
measurements, and the influence of compact object lensing in 
a ``clumpy Universe'' picture.

\section{Lensing magnification distributions} \label{sec.updf}

Consider a large number of perfect standard candles (sources with 
identical luminosities), 
distributed randomly on the sky at the same high
redshift $z$.  When observed in an inhomogeneous Universe,
they will possess a distribution of apparent luminosities, or 
magnitudes, due to alteration of the light 
ray bundle---gravitational lensing---by intervening structures along the
various lines-of-sight. The form of these lensing amplification
probability distribution functions (PDFs) depends both upon
the underlying cosmology ($\Omega_m$, $\Omega_w$, dark energy 
equation of state ratio $w(z)$, etc.) and upon the
nature of the structures causing the lensing (parametrized
by cosmological quantities such as the mass clustering amplitude 
$\sigma_8$, as well as 
details of the matter distributions: the mass function of
galaxies, specific density profiles of galaxies and clusters,
etc.). 

To determine the lensing amplification PDF we utilize the 
Stochastic Universe Method (SUM) presented in
\citet{hw98}.  This method is 
based upon a careful analysis
of the assumptions needed for a physically reasonable model
of a globally Robertson-Walker, but locally 
inhomogeneous, Universe. It 
calculates statistics for the full (weak and strong) lensing
probability
distribution of the observed image brightnesses of standard
candle sources at any redshift, for a variety of
cosmological models and a variety of mass
distributions. It does this in a Monte Carlo fashion,
randomly generating a portion of a Universe near a given
photon line-of-sight, and calculating the lensing effects
in this corner of the Universe. Repeating this many times, the full
lensing distributions can be well approximated. 

Comparisons of SUM to conventional N-body approaches can be found in
\citet{whm02}, which also argues that the lensing
probability distributions have a universal form. This
implies that fine details
of the distributions (e.g., those dependent upon unknown aspects of the
matter profiles of the lenses) will have an inconsequential
effect on broad characteristics of the PDFs, such as 
the variance, or the shift of the mode. As the current
analysis is dominated by these broad features, fine details
of the distributions will have little impact on our results.
We use the SUM approach, rather than the universal (weak
lensing) form presented in \citet{whm02}, to ensure that
any effects due to the strong-lensing tail are properly
modeled.
Unless otherwise noted, all the results which follow are for
a flat $\Lambda$CDM model, with $\Omega_m=0.28$. 
The dark matter is taken to be smoothly
distributed in halos (with density given by singular
isothermal spheres or Navarro-Frenk-White profiles; details
of the density profiles do not qualitatively change the
results \citep{hw98}). The further clumping of the dark matter into
compact objects (e.g., MACHOs) will enhance the effects of
lensing \citep{amanullah} (see also \citet{goobarsl} for the 
strong lensing case).

The general features of a lensing 
probability distribution function, $P(\mu)$, for
observed flux $\mu$, are as follows: $P(\mu)$ is 
peaked at a de-magnified value (compared to the ``unlensed''
pure Robertson-Walker result, which is normalized to $\mu =
1$), with a long tail to high magnification. The mean of the
distribution matches the Robertson-Walker value, $\langle\mu\rangle=1$, 
due to conservation of photon number. As the mode
is unequal to the mean, the distributions are manifestly
non-Gaussian.

Most sources will be slightly demagnified by the presence 
of structure, making them appear fainter and hence further away. If 
lensing were overlooked the resulting cosmography would be distorted, 
artificially enhancing the accelerating expansion caused by dark energy.  
This has 
a very small effect at low redshifts, however, since there is little optical 
depth to lensing (it is smaller by an order of magnitude than the detected 
acceleration~\citep{holz98}).  At high redshifts it would continue to bias toward
acceleration, diminishing the expected deceleration 
from matter domination.

The faint demagnification of the 
majority of sources is compensated for by a small number of
highly magnified sources. For sufficient statistics
this high magnification tail is well sampled, and the mean of the
distribution approaches the true source flux. 
Precision cosmology requires just such multiple measurements, 
automatically leading to the lensing distribution becoming better
sampled and less biased, 
whether we explicitly recognize the presence of lensing or not. 
For high enough statistics, therefore, lensing magnification
will not upset our 
interpretation of the acceleration and deceleration of the cosmic 
expansion, merely add noise to the precision measurements of its 
detailed behavior and the cosmological parameters derived
therefrom.

It is important to note that the 
lensing noise can be a signal in its own right.  The statistical 
characteristics of the dispersion carry information about the matter 
power spectrum, e.g.\ $\sigma_8$ \citep{friemanholzknox}, as
well as elucidating the nature of the dark matter \citep{ms99,sh99}.  The more
highly 
magnified sources can be useful cosmological probes as well, sensitive 
to both the cosmography and the large scale structure properties 
\citep{kochanek,kuhlen,hutererma,linsl}; 
in particular, calibrated candle sources such as supernovae allow 
greatly improved separation of the cosmological background information 
and the lens model \citep{holz01,oguri}.

This paper concerns itself with the impact of lensing on
determination of the distance-redshift relation, and the derived cosmological 
model.  We quantify the meaning of sufficient statistics to moderate the 
two effects: the non-Gaussianity, in the form of 
the shift of the mode from the mean, causing a
bias in the measured distance, and the dispersion of
the lensing distribution, contributing an additional source of
noise to the measurement. \S\ref{sec.dzperfect} addresses the case 
of standard (or perfectly calibrated) candles, while~\S\ref{sec.dzreal} 
treats the case of sources 
with some dispersion in their intrinsic luminosity. \S\ref{sec.cospar} 
then discusses the implications for cosmological parameter determination.

\section{Safety in Numbers}
\label{sec.convolve}

\subsection{Perfect Standard Candles}\label{sec.dzperfect} 

We begin with an idealized experiment: a large sample of
perfect standard candles at high redshift. Without lensing,
these would all be observed to have the same brightness: a
delta function PDF (normalized at $\mu=1$). We now add in the effects of
gravitational lensing, which both contributes a width to
the observed PDF, and shifts the mode of the PDF to slightly
demagnified values.
As already emphasized, the non-Gaussian lensing PDF
preserves the mean:
\beq
\left\langle{\mu}\right\rangle=\int\!\!d\mu\,\mu P(\mu) = 1.
\eeq 
This crucial property implies that, for
sufficiently high numbers of observed high-redshift standard
candles at a given $z$, the average brightness (in flux)
will converge to the
appropriate, unlensed brightness. It is to be noted,
however, that the second moment of the lensing PDF doesn't
necessarily converge.
For the case of point-mass lenses,
the probability at high magnification falls
off as $1/\mu^3$, and so the contribution
to the second moment is given by:
\beq
\left\langle{\mu^2}\right\rangle = \int\!\!d\mu\,\mu^2
P(\mu) \propto \int d\mu/\mu, 
\eeq
which diverges
logarithmically at high magnification,\footnote{In practice
this is mitigated by effects such as finite source size and 
obscuration.
If the second moment does diverge, then the
distribution of observed brightnesses of large numbers of
standard candles does not necessarily converge to a
Gaussian distribution by the central limit theorem, and statistical intuition based
upon normal statistics could lead us astray.} emphasizing
the non-Gaussian nature of the PDFs.

It is to be expected that the effects of non-Gaussianity
will be mitigated by observing sufficient numbers of SNe,
and more fully sampling the lensing PDFs.
With this in mind, we define $P_N(\mu)$ as the lensing
magnification PDF for the mean magnification of a sample of $N$ 
standard candles (at a fixed redshift). We
calculate $P_1(\mu)$ via the SUM code. The distribution for
higher numbers of standard candles can then be calculated
recursively: 
\beqa
P_N(\mu) &=& \int\!\!\!\int\!\!d\bar\mu\,d\bar\mu'\,P_{N-1}(\bar\mu)P_1(\bar\mu')
\delta\!\left({\mu-{(N-1)\bar\mu+\bar\mu'\over N}}\right), \\ 
&=& N\int d\bar\mu\,P_{N-1}(\bar\mu)\,P_1(N\mu-(N-1)\bar\mu). \label{eq.PN} 
\eeqa 
This recursion becomes 
particularly straightforward using spectral methods.
Defining $\tilde P_1(k)$ 
as the Fourier transform of the lensing PDF $P_1(\mu_1)$, the convolution 
becomes 
\beq 
P_N(\mu)=(2\pi)^{(N-2)/2}N\int dk\,\tilde P_1^N(k)\,e^{-ikN\mu}. 
\label{eq.PNk}
\eeq 
As expected, the convolution (eq.~\ref{eq.PN} 
or~\ref{eq.PNk}) preserves the normalization and 
mean of the distribution, and the variance
shrinks as $1/N$. 
This formula reduces to a simple expression in the case of 
Gaussian or log-normal probabilities (see Appendix \ref{sec.log}).

\begin{figure}[t]
\plotone{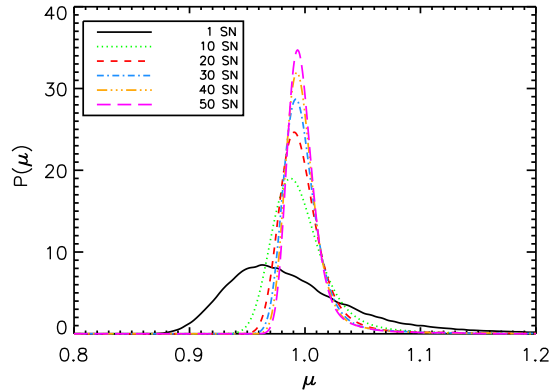}
\caption{Effective lensing magnification distributions for
multiple perfect standard candles, at $z=1.5$ in a
$\Lambda$CDM concordance cosmology. As more sources are
observed the distribution approaches a Gaussian, and eventually 
converges on a $\delta$-function
at the unlensed magnification, $\mu=1$.}
\label{f:pdf1}
\end{figure}

\begin{figure}
\includegraphics[scale = 0.5]{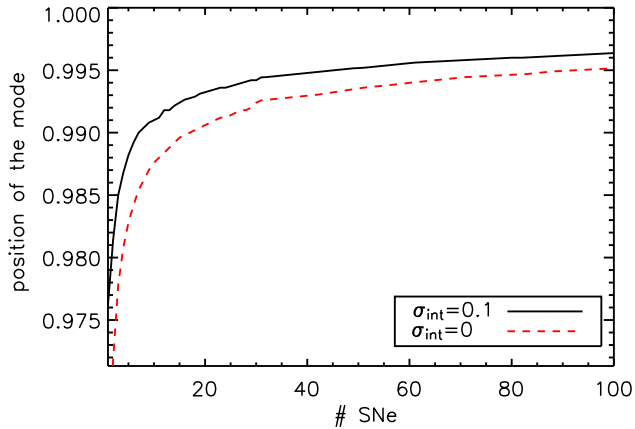}
\caption{Mode (peak) of the lensing magnification
distributions for multiple standard candles (see
Fig.~\ref{f:pdf1}), as a function of the number of SNe
averaged together. For a Gaussian, the mode equals the
mean. The (very slow) convergence to the mean (at 1) for many sources
averaged together can be seen. This is shown both for
perfect standard candles ($\sigma_{\rm int}=0$), and for SNe
with intrinsic noise given by a Gaussian in flux with
$\sigma_{\rm int}=0.1$. This is for SNe at $z=1.5$.}
\label{f:mode_N}
\end{figure}

The distributions at $z=1.5$, for various values of $N$,
are shown in
Figure~\ref{f:pdf1}.
Note that even for as many as 50 SNe averaged together, the
resulting distribution in 
magnification is still visibly asymmetric. This is a result
of the high-magnification tail possessed by the
lensing distributions. Figure~\ref{f:mode_N} displays the
shift of the mode of the distribution, as increasing 
numbers of SNe are observed. As expected, the curve
asymptotes (slowly) to a value of 1, since the mean of the
PDFs is always preserved, and for large numbers of SNe the PDF
should be well sampled.

Figures~\ref{f:sig_N} and~\ref{f:sig_N_log}
display the variance of the multiple SN lensing 
PDFs, as a
function of the number of SNe.
Since we are restricting our attention to smoothly clustered dark
matter (as opposed to point masses like MACHOs; see \citet{amanullah} 
for a treatment of that case), the variance
for $P_1(\mu)$ remains well-defined and finite.
The difference between $\sigma$ and the full width at 
half maximum (FWHM) determination of the width is further evidence of the
non-Gaussianity of the underlying lensing distribution: 
for a perfect Gaussian, $\sigma={\rm FWHM}/2.36$.
For $N=1$, the 
standard deviation is 1.61 times the 
FWHM/2.36. 
By $N=50$, this factor has gone down to 1.35, 
indicating a more Gaussian-like distribution.

\subsection{Sources with intrinsic luminosity dispersion}\label{sec.dzreal} 

If the sources were perfectly calibrated candles, then the distribution 
of observed fluxes would exactly mirror those shown in
Figure~\ref{f:pdf1} (i.e., reflect only the lensing
magnification suffered during propagation). 
However, astronomical sources, even calibrated candles such as
Type Ia supernovae, retain some intrinsic variation in their
luminosity. 
This gives an innate {\it a priori}\/ fuzziness in the distance-redshift relation,  
and the observed relation is thus a convolution of both the
intrinsic and lensing flux distributions.
In other words, the distribution of observed flux, $F$, is given by 
\beqa 
P(F) &=& \int dF_0\int d\mu\, p^{\rm int}(F_0)\,p^{\rm lens}(\mu)\, 
\delta(F-F_0\mu) \\ 
 &=& \int {d\mu\over \mu}\,\,p^{\rm int}\!\left(\frac{F}{\mu}\right) 
p^{\rm lens}(\mu), \label{eq.convolve}
\eeqa 
where $F_0$ is the intrinsic source flux.  
In terms of magnitudes, the PDF for a single lensed,
imperfect standard candle is given by
\beq 
P^{\rm int+lens}(m)=\int d\mu\,P^{\rm lens}(\mu)\,P^{\rm int}(m+2.5\log\mu). 
\eeq 
To form the N-fold convolution of N lensed, imperfect standard 
candle SNe, one takes the lensing + intrinsic, single SN PDF
of equation~(\ref{eq.convolve}) as $P_1$, and plugs this into equation\
(\ref{eq.PN}) or (\ref{eq.PNk}).
As before, the normalization and mean of $P_1$ are preserved 
for $P_N$.  

\begin{figure}
\includegraphics[scale = 0.5]{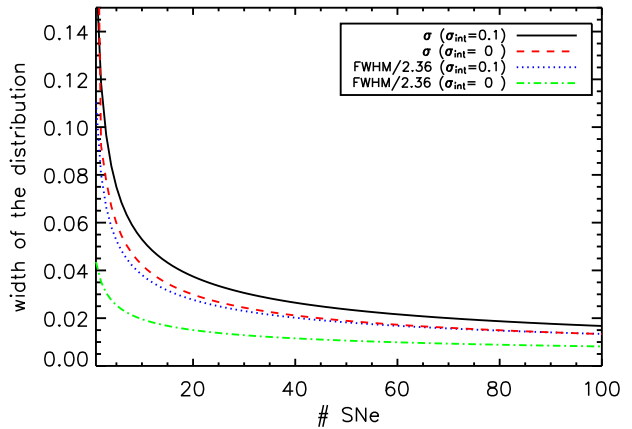}
\caption{Width of the lensing magnification
distributions for multiple standard candles (see 
Fig.~\ref{f:pdf1}), as a function of the number of SNe
averaged together. This is computed both as a
variance, $\sigma^2$, and as a full-width half-maximum,
FWHM. For a Gaussian, one has $\sigma=\mbox{FWHM}/2.36$. The
gap between the $\sigma$ and FWHM curves is an indication of the
non-Gaussianity of the lensing distributions. This is shown
both for perfect standard candles ($\sigma_{\rm int}=0$), and for SNe with
intrinsic noise given by a Gaussian in flux with $\sigma_{\rm
int}=0.1$. This is for SNe at $z=1.5$.}
\label{f:sig_N}
\end{figure}

\begin{figure}
\includegraphics[scale = 0.5]{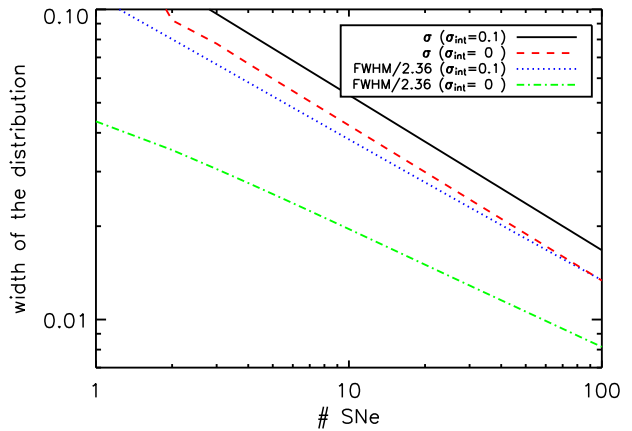}
\caption{As in Fig.\ \ref{f:sig_N}, but on a log-log plot.  The slopes 
of the standard deviation curves are very close to $-1/2$, indicating 
that the variance of the magnification distribution, including lensing, 
scales as $1/N$.} 
\label{f:sig_N_log}
\end{figure}

For the intrinsic supernova flux distribution we consider
both a Gaussian in 
magnitude (which corresponds to a log normal in flux) and a
Gaussian in flux. The SN distribution is characterized by a mean 
intrinsic magnitude ($\left<m(z)\right>=m_0(z)$ or
$\left<\mu(z)\right>=1$), and an intrinsic dispersion, $\sigma_{\rm int}(z)$.
Canonically one uses $\sigma_{\rm int}(z)=0.1$ or $0.15$
(constant with redshift).
Figures~\ref{f:mode_N}--\ref{f:sig_N_log} show the resulting
mode and variance, as a function of the number of observed
SNe, for intrinsic noise given by a Gaussian in flux with
standard deviation given by $\sigma_{\rm int}=0.1$.
Note that the addition of intrinsic noise to the sources
decreases the shift in the mode of the magnification PDF,
but increases the variance.

\subsection{Lensing as a function of redshift}
\label{sec.redshift}

The previous section concerned itself with the distribution
of the observed flux of multiple supernovae, averaged
together. These distributions are of interest because they
emphasize that lensing preserves the
mean (unlensed) flux, and thus for sufficient
numbers of observed SNe the lensing can be averaged
away. Figure~\ref{f:pdf1} is a visual representation of the process
of convergence to successively narrower Gaussians,
approaching the delta-function limit. 

Guided by this
convergence, we define an {\em effective
variance}, $\sigma_{\rm eff}^2$, due to lensing. Consider some
large number of standard candles, $N$, observed at a given
redshift. For sufficiently high $N$, the lensing
distribution is found to be well-approximated by a
Gaussian. If we designate the variance of this Gaussian as
$\sigma_N^2$, we can then define the effective variance due to
lensing, for a {\em single}\/ standard candle, as $\sigma_{\rm
eff}^2=N\sigma_N^2$. Although the true distribution of a single
standard candle is {\em not}\/ well-described by a Gaussian 
of standard deviation $\sigma_{\rm eff}$,
Figure~\ref{f:sig_N_log} indicates that the scaling of the
variance $\sigma_N^2$ 
is roughly inverse linear in $N$. So $\sigma_{\rm eff}$ 
gives 
an equivalent dispersion for each source when considering sufficiently
large, but finite, numbers of standard candles.  
A useful rule of thumb
for a reasonable use of $\sigma_{\rm eff}$ is to require 
$N$ greater than some lower limit calculated 
by demanding that $\sim$95\% of the
probability of the $N$-folded PDF be within $2\sigma_N$;
this yields values of $N\gtrsim10$. 

As SNe do not 
all occur at exactly the same redshift, we need to know over what 
redshift width we can apply this criterion.  As we take SNe 
spread out in redshift, the lensing PDFs, and hence $\sigma_{\rm eff}$, vary.
If we bound this variation such that the 95\% 
criterion above only dilutes to a $95\pm1$\% criterion, this is 
equivalent to using a $2\pm0.1\,\sigma_N$ allowance.  So we may 
group SNe within a 5\% fractional deviation in $\sigma_N$; since 
as we will see below, the standard deviation is linearly 
proportional to redshift, this imposes $\Delta
z/z\lesssim0.05$.
In other words, so long as there are $N\gtrsim10$
SNe in each redshift bin of width $0.1z$, the lensing can be
effectively described by a Gaussian with standard deviation
$\sigma_{\rm eff}$. (We emphasize that none of this necessitates any actual binning for 
the SN analysis.) For SNe that are more sparse
in redshift, the full non-Gaussian lensing PDF should be
used.
Sufficient numbers allow lensing to be treated as a scaled noise
added in quadrature; greater numbers of SNe are required to compensate
for this noise---see~\S\ref{sec.cospar}.  

Lensing has little impact at low redshifts, where a short
propagation distance implies a low optical depth. As the
distance to the source increases, the lensing effects become more
substantial. We thus expect $\sigma_{\rm eff}$ to increase with
redshift. Figure~\ref{f:sig_eff_z} plots 
$\sigma_{\rm eff}$ as a function of $z$, 
for perfect
standard candles, and for supernovae with intrinsic noise
given by a Gaussian in flux with $\sigma_{\rm int}=0.1$. 
The results can be well approximated by a linear fit: 
\begin{equation}
\sigma_{\rm eff} = 0.088z, \label{eq.sigeffz} 
\end{equation}
for the case of perfect standard candles. 
In terms of magnitudes, we find 
\begin{equation}
\sigma_{\rm eff,m} = 0.093z \label{eq.sigeffzm}
\end{equation}
(to be used if taking $\sigma_{\rm int}$ as 
Gaussian in magnitude).

\section{Cosmological Parameter Estimation in the Presence of 
Lensing} \label{sec.cospar}

\begin{figure}
\epsscale{1.1}
\plotone{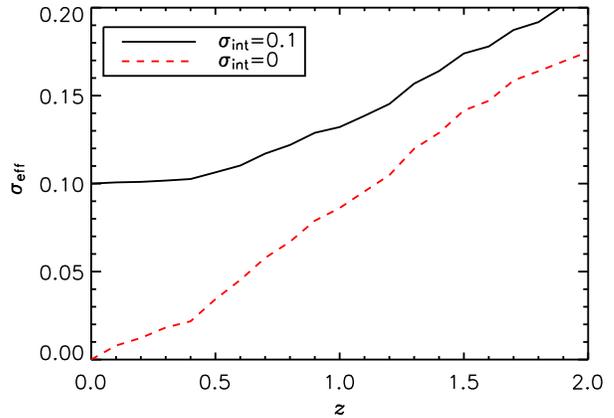}
\caption{The effective standard deviation in flux 
(see text) is plotted as a function 
of redshift, including lensing. For perfect standard candles
(dashed (red) line) this illustrates the pure lensing contribution; 
also plotted is the convolution with SN intrinsic dispersion given by
a Gaussian with standard deviation 0.1 in flux (solid (black) line). 
}
\label{f:sig_eff_z}
\end{figure}

To map the expansion history of the 
Universe and reveal properties of the dark energy, next generation 
experiments are being designed to 
observe the luminosity distance-redshift curve to 
one percent precision out to high redshift; for example, this is a 
primary science goal of the Supernova/Acceleration Probe\footnote{http://snap.lbl.gov}
\citep[SNAP;][]{snap}. 
From the cosmological inverse square law the luminosity distance, $d_L$,
is related to the flux magnification, $\mu$, by $\mu\sim
d_L^{-2}$. The errors on distance are then given by
$\delta d_L/d_L=(-1/2)\delta\mu/\mu\sim \sigma_\mu/2$.
It follows from~\S\ref{sec.convolve} (e.g., see
Fig.~\ref{f:sig_eff_z}) that the distance dispersion due to gravitational
lensing of any given SN at $z\ga0.5$ is of order or greater than
one percent.
Figures~\ref{f:pdf1}--\ref{f:sig_N_log} illustrate 
the way to overcome this noise: observations of sufficient
numbers of SNe per redshift ``bin'' convert a
high-dispersion, non-Gaussian lensing distribution to a
narrow, Gaussian-like one.  For example,  according to
Figure~\ref{f:sig_N}, in order for lensing to contribute
less than  
1\% distance uncertainty at $z=1.5$ requires $\sim70$ SNe (with
$\sigma_{\rm int}=0.1$), as compared to $25$ SNe if there
were no lensing.

It is to be noted that, since gravitational lensing is
achromatic, the redshifts of the sources are
unaffected. Thus the effects of lensing are confined to
changing the inferred luminosity distance.\footnote{Lensing
amplification can also alter which sources enter a flux
limited survey (Malmquist bias). However, most surveys are
designed to be forgiving of small fluctuations in the flux
threshold.} This compromises
the determination of the luminosity distance-redshift curve,
and thus impacts the estimation of cosmological
parameters. 

\subsection{Fisher approximation} \label{sec.fisher} 

Given an observable quantity, and a model for its
measurement errors, the most 
straightforward method for generating constraints on dependent cosmological 
parameters is the Fisher matrix approach. 
For a supernova survey, the data 
consists of magnitudes at several redshifts, and the Fisher matrix is
\beq 
F_{ij}=\sum_z \frac{\partial\lang m\rang (z)}{\partial\theta_i} 
\frac{\partial\lang m\rang (z)}{\partial\theta_j} \frac{N(z)}{\sigma_m^2(z)}, 
\eeq 
where $N(z)$ is the number of supernovae in a redshift
bin. 
This formalism can treat both offset effects in the observed magnitudes 
$\lang m(z)\rang$ and dispersion effects in $\sigma_m$. 

We take as a set of cosmological parameters $\theta_i=\{{\mathcal M},\om, 
w_0,w_a\}$, where ${\mathcal M}$ is a nuisance parameter involving the 
absolute magnitude of the supernova and the absolute distance scale 
(i.e.\ the Hubble constant), $\om$ is the dimensionless matter density 
(we assume a flat Universe), and $w_0$ and $w_a$ parametrize the dark 
energy equation of state ratio $w(z)=w_0+w_a(1-a)$.  

As a fiducial cosmology we adopt $\om=0.28$, $w_0=-1$, 
and $w_a=0$, and either include a Gaussian prior of $\sigma(\om)=0.03$ or 
a WMAP cosmic microwave background constraint on the distance to the 
last scattering surface. We also explore variations of the fiducial 
model. 

The increased magnification dispersion from lensing, as a function of 
redshift, is shown in Figure~\ref{f:sig_eff_z}, and given in the fitting 
function of equation~(\ref{eq.sigeffzm}).  
The total dispersion $\sigma_m(z)$ is then obtained by adding in quadrature 
the intrinsic SN luminosity dispersion and the lensing effect: 
\beq 
\sigma_m^2(z)=\sigma_{{\rm eff},m}^2(z)+\sigma_{\rm int}^2. \label{eq.sigm2}
\eeq 
One can verify from Figure~\ref{f:sig_eff_z} that the fitting form convolved 
with the intrinsic dispersion is a good approximation to
the full, numerical solution. 

Before we study the effect of the increased dispersion on determination 
of cosmological parameters, we obtain a first indication of the influence 
of lensing by asking how 
many supernovae are required to offset the increased dispersion. 
We can then correct for this 
extra scatter by increasing the sample size. 
If we want the magnitude scatter to be at 
the same level as if there were no lensing, then we observe extra supernovae 
such that 
\beq 
N(z)=N_0(z)\,\sigma_m^2(z)/\sigma_{\rm int}^2. \label{eq:ndegrade} 
\eeq 

\begin{figure}
\epsscale{1.1}
\plotone{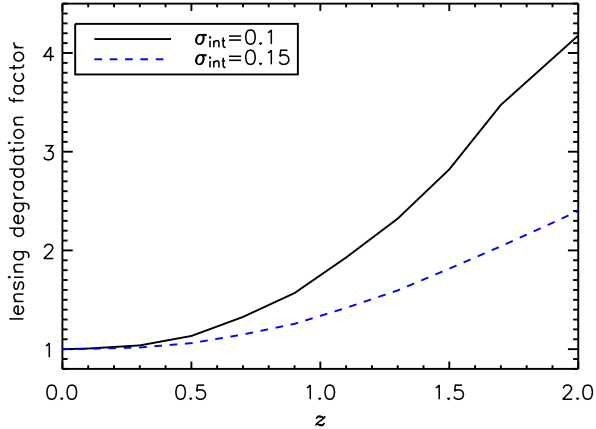}
\caption{Degradation factor due to gravitational lensing, as
a function of redshift. This gives the
fractional increase in the number of standard candles to be
observed, to reduce the dispersion to what would have been
observed without lensing. This is shown for two values of
the SN intrinsic noise, given by Gaussians in flux.
}
\label{f:degradation.z}
\end{figure}

From equations~(\ref{eq.sigeffzm}) and (\ref{eq.sigm2}) with
$\sigma_{\rm int}=0.1$, 
0.15, we find that the excess numbers required are $\Delta N/N_0=(z/z_*)^2$ 
with $z_*=1.08$, 1.61 respectively (i.e., in the bin around $z_*$ half of 
the SNe are simply going to offset the increased lensing dispersion). 
Figure~\ref{f:degradation.z} illustrates this ``survey
bloat'', or conversely 
the degradation factor in the effectiveness of $N$ SNe. 
At $z=1.5$ one requires nearly three times 
as many SNe (with intrinsic dispersion $\sigma_{\rm int}=0.1$
in flux) to obtain the same dispersion 
as the case without lensing. In other words, each $z=1.5$ SN
has $1/3$ the statistical weight it would have had in the
absence of gravitational lensing.

For sufficient number of SNe (given by the degradation
factor in eq.~\ref{eq:ndegrade}), however, lensing dispersion has no ill 
effect on cosmological parameter estimation.  Furthermore, because the 
cosmological leverage occurs not merely from the highest redshift SNe, but 
from SNe at all intermediate redshifts as well, the relevant 
degradation factor 
is much more modest than suggested in Figure~\ref{f:degradation.z}. 
For a fiducial 
distribution of 40 low redshift SNe, 10 per 0.1 bin at
medium redshift, and 1 per 0.1 bin 
at high redshift (with no systematic uncertainties other than lensing), 
and adding a 0.03 prior on $\om$, the effect of lensing 
from equation~(\ref{eq.sigm2}) is a 15--20\% weakening of the constraints on 
$w_0$ and $w_a$.  Furthermore, this can be undone by increasing 
uniformly the number of SNe in each bin by a factor of 1.4, not 
the factor of three suggested by Figure~\ref{f:degradation.z} for 
$z=1.5$ alone.  (These numbers hold with the further
addition of a WMAP prior 
on distance to the last scattering surface.  For a fiducial cosmology 
with SUGRA dynamical dark energy ($w_0=-0.82$ and $w_a=0.58$), rather than a 
cosmological constant, the weakening is 20--30\%, though now a WMAP 
prior helps reduce it to 5--7\%.) 

\begin{figure}
\epsscale{1.1}
\plotone{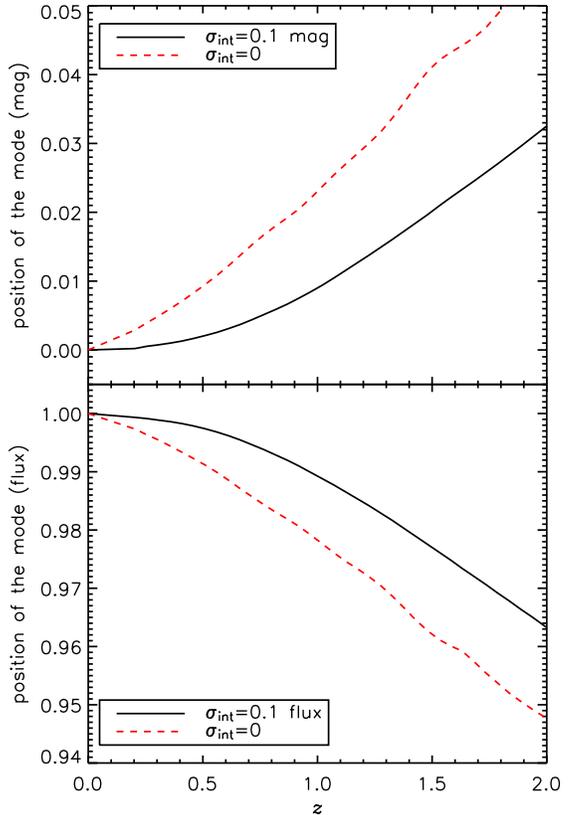}
\caption{The mode of the distributions, as a function of
redshift. This is shown both in magnitude (upper panel) and
flux (lower panel), for either SN intrinsic noise (in
magnitude or flux, respectively),
$\sigma_{\rm int}=0.1$ (solid black), or perfect standard
candles (dashed red).
}
\label{f:mode_z}
\end{figure}

In addition to dispersion, lensing leads to a skewness in 
the convolved flux distribution.  For a large sample this is not 
worrisome since lensing preserves the mean flux; a proper SN cosmology 
analysis treats the data in flux (though it may be quoted in terms of 
magnitude), and so there is no offset in the mean (see Appendix \ref{sec.log} 
for a discussion of bias induced by flux-magnitude 
confusion).  But if 
the distribution is not well sampled then one might find a magnification 
offset from the mean.  We characterize this by the mode of the 
distribution, which will be shifted to magnification lower than unity.  
Such an offset would 
affect cosmological parameter determination, and we can calculate this 
within the Fisher matrix formalism.  
Figure~\ref{f:mode_z} shows the mode offset, reasonably fit 
for a SN with $\sigma_{\rm int}=0.1\mbox{ mag}$ by
\beq
\Delta m(z)=0.025\,(z-0.7),\qquad z>0.7 \label{eq.mode}
\eeq
and zero shift for $z<0.7$.

Such an offset propagates into a cosmological parameter bias via
\beq
\Delta\theta_i=(F^{-1})_{ij}\int dz\,N(z)\,\Delta m(z)\frac{\partial
m}{\partial \theta_j}\frac{1}{\sigma_m^2(z)},
\label{eq.bias}
\eeq
and gives a bias of less than $0.2\sigma$ (for the SUGRA case as well), 
where $\sigma$ is the 
statistical error. Indeed, the shift of the mode is of even less 
concern in that
equation~(\ref{eq.mode}) refers to a single SN.  For $N$ SNe at a
given redshift, we expect the magnification distribution to
approach a Gaussian, and hence the mode to approach unity, 
with its deviation diminishing as $1/\sqrt{N}$. 
This behavior is demonstrated in 
Figure~\ref{f:mode_N}. Since the
Fisher matrix increases proportionally with $N$ (it is also known as the
information matrix), then from equation~(\ref{eq.bias}) we see that the
parameter bias will scale as the mode offset, and thus
vanishes as $N^{-1/2}$.  For the number of SNe in current
data sets, 
the lensing bias makes little difference (unlike the flux-magnitude 
asymmetry; see Appendix \ref{sec.log}) due to the large statistical 
error, but as the number of SNe increases the scaling will make the 
mode offset effect essentially negligible. 

The overall conclusion of the Fisher analysis is that lensing is a 
relatively benign error.  It is of significance only at high
redshifts ($z>1$), 
where there are currently few SNe measurements, and so large statistical 
errors, and hence only modest cosmological leverage.  Once
many high-$z$ SNe are observed, however,
then lensing again is not a major source of error since the 
mean magnification averages to unity.  Only when there are merely a 
handful of high redshift SNe per bin will lensing play a significant role.

\subsection{Monte-Carlo simulations} \label{sec.monte-carlo}

While the Fisher analysis of the previous subsection gives a 
simple, intuitive feel for the effect of lensing on cosmological 
parameter determination, there remains the issue of 
the non-Gaussian nature of the gravitational lensing signal. 
A more robust, if computationally intensive, 
approach is to directly simulate data sets, and explore the
ensuing scatter in best-fit parameter space. This presumes
that the likelihood of a set of derived parameters, given an
observed data set, is equivalent to the probability of
observing a data set, for a given set of parameters.

We specify a cosmology, and calculate the lensing
distribution via the SUM approach. We then simulate an
``observed'' data set, with the SNe at pre-defined
redshifts, and the intrinsic and lensing noise for each SN
drawn randomly from the appropriate distribution. For each
data set we find the best-fit cosmology (using a maximum
likelihood approach). We repeat this procedure for many
($\sim$40,000) simulated data sets, and draw contours in
parameter space encompassing the distribution of
best-fit models (e.g.\ the $1\sigma$
contour is the smallest contour which encompasses $68\%$ of
the simulated data sets).

\begin{figure}
\epsscale{1.1}
\plotone{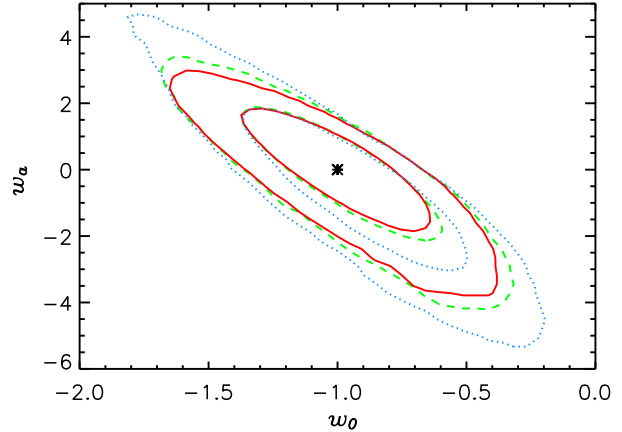}
\caption{Comparison of contours in parameter space, both
with and without lensing. This is for observations of 20 SNe evenly
distributed in the range $0.01<z<0.1$, 20 SNe in
$0.1<z<0.5$, 8 SNe in $0.5<z<0.9$, and 8 SNe in $1<z<1.7$.
The intrinsic dispersion of the supernovae is taken to be a
Gaussian in magnitude with standard deviation given by
$\sigma_{\rm int}=0.1\mbox{ mag}$, with a prior
on $\Omega_m$ given by $\Omega_m=0.28\pm0.03$. This data set
is roughly comparable to the current state of SN
observations (which are statistically more numerous, but with greater
individual errors).
Solid (red) contours are without lensing. Dashed (green) contours include
lensing effects. Dotted (blue) contours represent an extreme lensing
scenario, where 1/4 of the dark matter is in compact objects
(MACHOs).
}
\label{f:contour1}
\end{figure}

Figure~\ref{f:contour1} shows the resulting $1\sigma$ and $2\sigma$ 
contours, for
a SN set roughly representing the current data (see the
caption for details). 
The presence of lensing makes only a
slight difference in the contours. This is because the data
set under consideration has a small number of
high-redshift ($z>1$) SNe, and so the effects of lensing are
mitigated.  At $3\sigma$, however, lensing has a more significant 
influence, greatly increasing the dispersion.  
In addition, lensing due to compact objects 
(MACHOs, e.g.\ solar-mass black holes), will have a greater 
impact.  We consider the extreme case of
$1/4$ of the dark matter in compact objects, representing 
the maximum reasonable lensing effects 
for a given cosmology. The outermost (dotted) set of contours
illustrate this extreme case; the additional lensing leads to
a $\sim30\%$ degradation of the parameter estimation (also see 
Appendix \ref{sec.apxclump} on clumpy lensing).

The moral to be
drawn from this figure is that lensing is unlikely to have
dramatic effects on the cosmological parameters estimated from
current supernova data.  A similar conclusion holds if one 
adopts a CMB prior rather than a matter density prior.  
However, we expect increasing influence of lensing as 
observations attain higher redshifts, but have not yet 
achieved sufficient numbers of supernovae to compensate for the lensing degradation.  

\begin{figure}
\epsscale{1.1}
\plotone{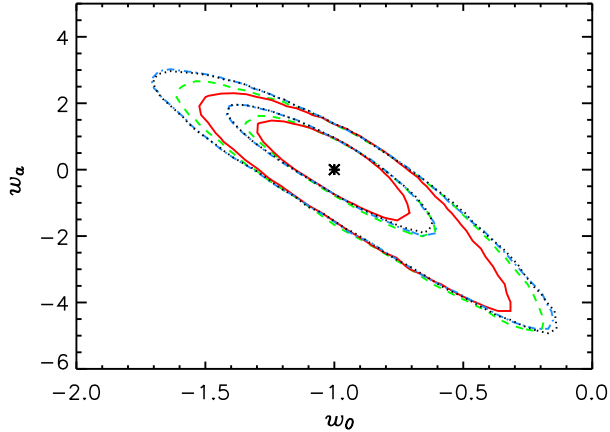}
\caption{At redshifts $z\ge1$, lensing effects become more significant. 
With sufficient numbers of supernovae, however, the effects can be 
treated via a Gaussian effective dispersion, or number degradation factor,
and well approximated with the Fisher matrix formalism.  Here contours 
represent parameter estimation from 40 SNe at $z<0.1$, 100 SNe at $z=1$, and 100 at $z=1.5$,
including a prior on $\Omega_m$ given by
$\Omega_m=0.28\pm0.03$. 
The solid (red) contours are for SNe with intrinsic
dispersion given by a
Gaussian in magnitude with standard deviation
$\sigma_{\rm int}=0.1\mbox{ mag}$, without any gravitational lensing
effects considered. The dashed (green) contours are for the same data
set, but with the effects of lensing included through
Monte Carlo of the PDF.
The dotted (black) contour represents the $\sigma_{\rm eff}$ case
(see~\S\ref{sec.redshift}), where the same SN data set is
considered, with the lensing effects approximated by
an increased Gaussian intrinsic dispersion in magnitude for
the SNe (see eq.~(\ref{eq.sigm2})).
The dot-dashed
(blue) contours demonstrate the degradation factor
(see~\S\ref{sec.fisher}). This is for SNe with intrinsic
noise given by $\sigma_{\rm int}=0.1\mbox{ mag}$, but with the
lensing approximated by a diminution in the number of SNe,
in accordance with eq.~(\ref{eq:ndegrade}), yielding 54 SNe
at $z=1$ and 34 SNe at $z=1.5$. As expected, the $\sigma_{\rm
eff}$ and ``degradation'' contours are essentially
indistinguishable.
}
\label{f:z1and1.5} 
\end{figure}

Figure~\ref{f:z1and1.5} shows that once high redshift surveys 
do observe many supernovae, the effective dispersion formula of 
equation~(\ref{eq.sigeffz}) provides good insight, and the Fisher 
matrix formalism, properly handled, can give a reasonable 
approximation.  With 100 SNe at $z=1$ and 100 at $z=1.5$, the 
contour including lensing is bloated by 30\% in $w_0$ and 20\% 
in $w_a$ relative to a Universe without lensing.  The 
Monte Carlo contour can be predicted by using 
the appropriate $\sigma_{\rm eff}$, or equivalently degraded number 
of SNe.

Specifically, we see that the lower right half of the Monte Carlo 
lensing contour (dashed (green) curves) matches the Monte Carlo contours generated without 
the presence of lensing, but with dispersion corresponding to 
$\sigma_{\rm eff}$ added in quadrature (dotted (black) curves), 
or with merely intrinsic dispersion 
but numbers degraded by the appropriate factor from 
equation~(\ref{eq:ndegrade}) (dot-dashed (blue) curves).  
The upper left half of the lensing 
contour is not as extended, and approaches a Monte Carlo contour 
generated without lensing (solid (red) curves).

We can understand this as follows: 
The models toward the lower right have equation of state ratios 
$w<-1$ at redshifts $z\gtrsim1$ (where lensing is important).  For such 
values of $w$, the SNe should be dimmer than for our fiducial model. 
However, lensing magnification, $\mu>1$, can offset this dimming and 
make these models of nearly equal likelihood to the fiducial---i.e.\ 
they will lie within the confidence level contour.  For models toward 
the upper left, on the other hand, the opposite obtains:
$w>-1$ and $\mu<1$.
The $\mu>1$ part of the lensing PDF, convolved with the 
intrinsic PDF, is well represented in its cosmological effects by 
a Gaussian with $\sigma_{\rm eff}$.  The $\mu<1$ part, even after 
convolution, is narrower due to the sharp low $\mu$ behavior of the 
lensing PDF, and so gives a Gaussian with dispersion closer to 
$\sigma_{\rm int}$.
This can be understood physically: gravitational lensing imposes a
lower limit to the amount of {\em de}\/magnification, given by the
empty-beam value (e.g., there is a minimum amount of
focusing due to matter along any
line-of-sight, given by the vacuum value). There is no
similar limit to the magnification side. Thus, lensing is by
its nature asymmetrical, and this is reflected in the contours.
Note, however,
that both sides can independently be reasonably approximated by the 
Fisher formalism with the appropriate dispersions.

\begin{figure}
\epsscale{1.1}
\plotone{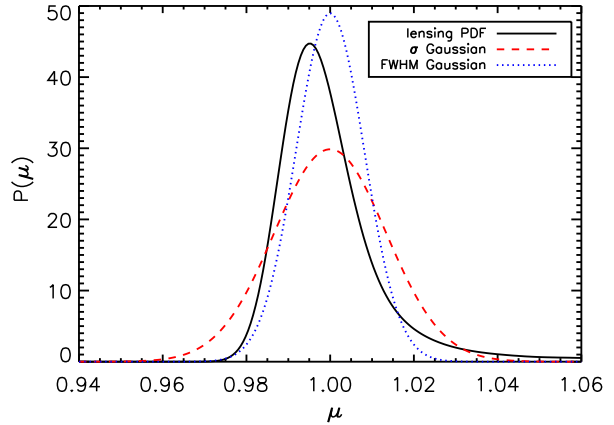}
\caption{When many high redshift sources are
observed, the lensing magnification distribution will approach a 
Gaussian and eventually converge on a $\delta$-function
at the unlensed magnification, $\mu=1$.  However, this convergence 
can be slow; the solid (black) curve shows the PDF for 100 SNe at $z=1.5$. 
Residual non-Gaussianities are clear: the lensing PDF shows
deviations from both the dashed (red)
curve representing a Gaussian with the same variance,  and
the dotted (blue)
Gaussian distribution with the same full width at half maximum.  
However, we find that these residuals do not substantially affect cosmological 
parameter estimation, and it can be sufficient to account for
lensing by means of a Gaussian effective dispersion $\sigma_{\rm eff}$. 
}
\label{f:notgaussian} 
\end{figure}

This general agreement is despite the fact that the PDF has not yet
converged to a Gaussian form.  Figure~\ref{f:notgaussian} clearly 
indicates that visible non-Gaussianity remains in the probability 
distribution even with 100 SNe at $z=1.5$: the lensing PDF is skewed from 
a Gaussian with the same variance (solid (black) curve vs.\
dashed (red) curve), and the variance is significantly different from 
the FWHM/2.36 (dotted (blue) curve is a Gaussian with the same FWHM 
as the lensing PDF).  Thus, both the mode and the width of the 
magnification distribution exhibit non-Gaussianity---the central 
limit theorem still has a ways to go.  Note, though, that
the mean values for all 
the distributions agree (at the (unlensed) value of 1), and Figure~\ref{f:z1and1.5} 
indicates that the cosmological parameter 
estimation is only dependent on broad characteristics of the 
lensing PDF, sufficiently described by our $\sigma_{\rm eff}$.

\section{Conclusion} \label{sec.concl}

Cosmological observations of the distance-redshift relation
of Type Ia supernovae have revealed an extraordinary new
component in the Universe: dark energy.  To explore this in
detail, we must comprehend the astrophysics that factors
into the measurements and their interpretation.  Realistic
assessment of the data from large surveys over the next
decade requires consideration of a realistic Universe.
Structure on various scales fills the Universe, and {\em
every} light ray is affected at some level by the resulting
gravitational lensing.

We have presented quantitative calculations of lensing effects on 
cosmological parameter estimation through observation of the luminosity 
distance-redshift relation.  Using a Monte Carlo simulation code 
we derived the lensing magnification 
probability distribution function, and investigated its non-Gaussian 
characteristics, its behavior as the number of sources increased, and its
redshift dependence.  All of these have important 
ramifications for the interpretation of high-redshift
observations. 

In particular, above redshift $z\simeq1$ lensing has a significant effect, 
degrading the statistical weight of sources by factors that exceed two 
(i.e.\ each source has less than $1/2$ the cosmological 
leverage of 
one in a Universe without lensing), and increasing roughly as $z^2$. 
Standard candles at redshifts $z\gtrsim2$ will be severely impacted 
unless several hundreds can be observed at each redshift; this seems to 
doom hypothetical ultra-high-redshift standard candles such as 
gamma ray bursts or 
gravitational waves from inspiraling black holes 
(see \citet{holzhughes03,holzhughes04} 
for a similar conclusion). 

We provide useful fitting formulas for the effective luminosity variance 
and mode offset induced by lensing, as a function of redshift, and 
compute their influences on cosmological parameter determination, 
checking these against Monte Carlo simulations.   Lensing effects, 
though still appreciably non-Gaussian even with 100 sources at a given 
redshift $z>1$, can be well approximated by fitting functions 
for $N\ga10$ sources (within $\sim0.1z$).
The fitting function for the flux dispersion induced by
lensing is given by:
\beq 
\sigma_{\rm eff}=0.088z.
\eeq 
From this expression, and from direct numerical
calculations, we find that $N\approx70$ 
sources are needed to achieve 1\% distance measurements at $z=1.5$. 
This effective dispersion formalism, combined with the flux conserving 
properties of lensing (the mean magnification is unity), tells us that 
while lensing certainly cannot 
be ignored, it is tractable, with safety in numbers.

\acknowledgments

We thank the Kavli Institute for Theoretical Physics, the Michigan 
Center for Theoretical Physics, and the Aspen Center for Physics for 
hospitality at various stages during this work. DEH has been
supported by NSF Grant PHY-0114422 to the KICP, and
gratefully acknowledges a Feynman Fellowship from LANL.
EVL has been supported in part by the Director, Office of Science, US Department 
of Energy under grant DE-AC03-76SF00098. 

\begin{appendix} 

\section{Non-Gaussianity: Log-normal approximation and flux vs.\ 
magnitude} \label{sec.log}

While the full lensing PDFs derived from Monte Carlo simulations 
within the Stochastic Universe Method encapsulate the effects of 
structure on the luminosity distance-redshift relation, it is also 
useful to consider a simple approximation exhibiting non-Gaussianity. 
This will allow derivation of a number of analytic relations and 
help provide insight. 

Since astronomical observations of supernovae use a logarithmic flux 
system of magnitudes, with $m=-2.5\log_{10} F$, it is natural to investigate 
a log normal distribution as a stand-in for the lensing magnification.  
This in fact has 
several attractive properties: a natural skewness, simple characterization, 
and Gaussian behavior near the peak. 
We do not claim it fits the derived lensing PDF, only that it has 
some related properties. 

A log normal probability distribution has the form
\beq 
p(\mu)=\frac{1}{\sqrt{2\pi}} \frac{1}{S\mu} e^{-(\ln\mu-U)^2/(2S^2)},
\eeq 
and possesses mean $\lang\mu\rang=e^{U+S^2/2}$, with mean 
logarithm $\lang\ln\mu\rang=U$, and variance $\sigma_\mu^2=\lang\mu^2
\rang-\lang\mu\rang^2=e^{S^2}-1$.  
Note that from the physical interpretation of the magnification, flux 
conservation ensures that $\lang\mu\rang=1$, and so for our case 
$U=-S^2/2$. Thus we have a one parameter PDF with $\lang\ln\mu\rang 
<0$, skewed negative 
as desired. 

We can now evaluate equation~(\ref{eq.convolve}) and find that the convolution 
distribution for the observed flux is also a log normal distribution, 
broader and with different skewness: 
\beq 
P(F)=\frac{1}{\sqrt{2\pi}} \frac{1}{\sqrt{S^2+\sigma_{\rm int}^2/b^2}} \frac{1}{F} 
e^{-(\ln F+m_0/b-U)^2/[2(S^2+\sigma_{\rm int}^2/b^2)]}, 
\eeq 
where $b=2.5/\ln 10$ converts from flux in natural log to
magnitudes.
The mean flux is $F_0$, and the mean of the 
logarithmic flux is
\beq 
\lang\ln F\rang=-\frac{\ln 10}{2.5} m_0+U = \lang\ln F_0\rang 
+\lang\ln\mu\rang. 
\eeq 

For observations using magnitudes, it is convenient to convert to the 
distribution in magnitude, $m$:
\beq 
P(m)=\frac{1}{\sqrt{2\pi}\,\sigma_m} e^{-(m-m_0+bU)^2/(2\sigma_m^2)}, 
\eeq 
where $\sigma_m^2(z)=\sigma_{\rm int}^2+b^2S^2$.  
Also, recall that from flux conservation $U=-S^2/2$.  Thus these 
probability distributions for source properties and lensing effects 
lead to a particularly simple result for the observed magnitude-redshift 
relation: there is an increased dispersion $\sigma_m$, and an offset in 
the mean.  Explicitly, 
\beqa 
\lang m\rang (z) &=& m_0(z)-bU(z) \label{eq.mean} \\  
\sigma_m^2(z) &=& \sigma_{\rm int}^2-2b^2U(z).  \label{eq.sigm} 
\eeqa 
Recall that $U<0$.  The increased dispersion will weaken 
cosmological parameter determination, and the offset in the mean 
will bias the derived parameters.  The magnitude of the effect is 
controlled by the quantity $U=\lang\ln\mu\rang$. 
For example, the variance is doubled, relative to a 0.1 mag 
intrinsic dispersion, for $U=0.004$. 

It is worth noting that no actual lensing is required to
generate bias in the cosmological parameter
determination. Image intensity can be expressed in either
flux or magnitude, and the transformation between these
systems of measurement gives a similar form of
non-Gaussianity---that of the log normal distribution---with 
the attendant dispersion and bias.  By
confusing flux and magnitude values in an analysis of SN
data, it is possible to severely compromise the values of
the resulting cosmological parameters.  Observing sources
given by the numbers and redshifts in~\citet{riess04}, with
an intrinsically Gaussian flux distribution, but
interpreting them in terms of magnitudes, turns a true flat
cosmology with $\Omega_m=0.28$ and a cosmological constant
into an apparently substantially closed cosmology
($\Omega_k\approx0.2$).  This might also be interpreted as a flat
model with a dark energy different from $w_0=-1$, $w_a=0$, in 
particular $w_0<-1$, $w_a>0$. 
To avoid bias one must therefore carry out both the
observations and the analysis in a fully consistent
manner~\citep{hlr04}.

\section{Clumpy Universe model} \label{sec.apxclump} 

The central problem of incorporating lensing effects from structure 
in the Universe into the distance-redshift relation is calculation 
of the magnification probability distribution function (see~\S\ref{sec.updf} 
of the main text).  
The simplest distribution of lensing magnifications,
however, is no distribution at all.  This is the approach of
the clumpy Universe model of \citet{dyerroeder}.  This model does not 
ignore lensing, but rather assumes a 
special set of lines of sight---those that avoid mass concentrations. 
This preferentially avoids magnifications greater than unity.  
To conserve mass the 
model partially evacuates most of space, leaving an average density 
$\alpha\rho$, where $\alpha$ is the smoothness parameter and $\rho$ the 
globally averaged energy density, and concentrates the remaining mass 
into compact clumps that the ``chosen'' light rays never encounter.  
Furthermore, 
shear effects from these inhomogeneities are ignored.  Thus the 
only light paths the observer deals with lead to demagnification ($\mu<1$) 
of the source, and overall flux is not conserved (we mention a way around 
this below). 

We characterize this as shunning lensing rather than ignoring it, 
willfully neglecting data from sources touched by magnification; in 
general this is an even worse idea than completely ignoring lensing, 
due to the bias and flux nonconservation.  However, we can use it as 
a simple, extreme case to examine the effect of lensing on the 
distance-redshift relation, and the ensuing cosmological parameter 
determination and bias (for example, see~\S\ref{sec.monte-carlo} and 
Fig.~\ref{f:contour1}).  Thus it is worth investigating some aspects 
of it here. 

The distance-redshift relation is no longer an integral over the Hubble 
parameter, but instead becomes a second order differential equation following from the 
optical scalar equations.  This was given for general equations of state 
by \citet{kayser} and \citet{lin88}, and can treat 
inhomogeneities not 
only in the matter but hypothetical clumpiness in the dark energy component. 
The effect of clumpiness enters the Ricci focusing 
as $[1+w(z)][1-\alpha_w(z)]\Omega_w(z)$, 
with $\alpha_w$ giving the ratio of smoothly distributed to total energy 
density in a component characterized by equation of state $w$.

We note several points: (1) a pure cosmological constant has no such Ricci 
focusing, (2) the effect is diminished as $w\to -1$ but is enhanced for 
larger $\Omega_w$, and (3) an evolving clumpiness can be treated 
straightforwardly (as detailed in \citet{lin88}).  One could certainly 
assume the canonical smoothness $\alpha_w=1$ for a component other 
than matter. 
However we can also explore more arcane models, such as postulating that 
dark energy couples to (dark) matter, or is affected through backreaction 
on the expansion from nonlinear structure formation in the matter. 
Thus we examine both $\alpha_w=1$ and $\alpha_w=\alpha_m$, where 
$\alpha_m$ is the clumpiness of the matter component.  Since the clumpy 
Universe model generally counts larger scale structure such as 
galaxies and clusters as a smooth 
component, only excluding compact objects with strong magnifications, 
we take the conservative position that all dark matter is smooth, and 
about half of the baryons as well, so $\alpha_m=0.9$.  The effects 
will be greater for more extreme cases.  If we take as a dark energy 
model the SUGRA model with $w_0=-0.82$ and $w_a=0.58$, then at 
$z=0$ the Ricci focusing due to clumpy dark energy would be roughly half 
that due to clumpy matter.  Continuing to be conservative, we henceforth 
ignore that contribution and assume the dark energy is smooth. 

Before investigating the influence of clumpy lensing on cosmological 
parameter determination, let us consider parameter degeneracies. 
Recently, \citet{caldwellkamion} pointed out that formally one could 
measure the curvature of space through the distance-redshift relation, 
with the curvature entering at third order in a low redshift expansion; 
however, the space geometry is entangled with the spacetime geometry, i.e.\ 
the expansion history.  This usefully highlights the same point made in 
\citet{lin88}.  Here we extend that result to show a six-fold degeneracy 
at ${\mathcal O}(z^3)$ of the angular diameter distance: 
\beqa 
r(z)\approx z-(1/2)(3+q_0)z^2+z^3[2+(\omt/4)(2+5\bar w-3\overline{w^2}-
\overline{w'} 
-\bar\alpha-\overline{w\alpha})+(1/2)q_0^2]+{\mathcal O}(z^4). 
\eeqa 
The notation $\bar X=\sum_w X\Omega_w(0)/\omt$, where $\omt$ is the total 
dimensionless energy density, related to the curvature density by 
$\Omega_k=1-\omt$ and to the deceleration parameter by $q_0=\omt 
(1+3\bar w)/2$.  Thus one must know the average of the equation of state 
ratio $w$, of the time variation $w'=dw/dz$, smoothness $\alpha$, etc. to 
determine the spatial curvature.  This points up the well 
known necessity for distance observations at many redshifts over a long 
baseline, e.g.\ $z\approx 0-2$, to break the degeneracies and determine 
the cosmological parameters. 

Let us consider such observations from a deep survey, corresponding to 
supernova observations from SNAP.  If we shun lensing, but
do not know 
the clumpiness of the Universe, then we introduce an additional 
fit parameter in the smoothness $\alpha$.  While this increases the 
uncertainties in the other parameters, the more insidious effect is 
bias, or misestimation of the parameters.  Within the context of a 
SUGRA model, assuming a smooth Universe while only utilizing lines of 
sight with $\alpha_m=0.9$ results in a bias to $w_0$ of $0.3\sigma$,
and to $w_a$ of $0.8\sigma$, where $\sigma$ is the statistical 
precision.  This gives a measure of the peril of shunning
lensing, and an estimate of the damage of low statistics.

We emphasize that homogeneous observations that do not select special 
lines of sight do not run into such bias---it is better to ignore 
lensing than shun it, and best of all is to acknowledge and include 
it as done in this article. 
Could we form a probability distribution such as 
simulated in the Stochastic Universe Model from an analytic convolution 
of clumpy lines of sight?  That is, could we incorporate both underdense 
and overdense lines of sight in such a way as to conserve flux and allow 
a simpler treatment than Monte Carlo analysis?  

Two difficulties exist with this method.  First, 
overdense lines of sight reach a limit at some critical $\alpha$, e.g.\ 
$\alpha>25/24$ for the flat, pure matter model, where the increased 
focusing forms caustics---i.e.\ the light rays focus before reaching the 
observer, and the angular distance vanishes there.  But this can be 
overcome by analytic continuation.  The second problem is that averaging 
must be done not in terms of density, i.e.\ restoring $\lang\alpha\rang=1$, 
but in terms of flux conservation $\lang r_\alpha^{-2}\rang=r_{FRW}^{-2}$, 
where FRW means a completely smooth, $\alpha=1$ Universe.  This requires 
the use of a redshift dependent smoothness parameter, which can be 
managed, but it also becomes a recursive relation (cf.\ \citet{lin88}, 
Appendix C), which is no easier to deal with than the physically 
motivated PDF formalism used in the text of this paper. 

Thus, our conclusions are that the Stochastic Universe Model, and 
the fitting functions we derive, offer the best way to realistically 
include lensing from the presence of structure in the Universe.

\end{appendix}

\end{document}